# Alternative formulations for gilthead seabream diets: towards a more sustainable production

Cláudia Aragão[1,2*], Miguel Cabano[1], Rita Colen[1], Juan Fuentes[1] and Jorge Dias[3]

*[1] Centro de Ciências do Mar do Algarve (CCMAR), Faro, Portugal*

*[2] Universidade do Algarve, Faro, Portugal*

*[3]SPAROS Lda., Olhão, Portugal*

*Author for correspondence:

Cláudia Aragão

E-mail address: caragao@ualg.pt

Centre of Marine Sciences (CCMAR)

University of Algarve,

Campus de Gambelas, building 7,

8005-139 Faro, Portugal

Tel: +351 289 800 100 x 7374

## Abstract

To support the expected increase in aquaculture production during the next years, a wider range of alternative ingredients to fishmeal is needed, towards contributing to an increase in production sustainability. This study aimed to test diets formulated with non-conventional feed ingredients on gilthead seabream (*Sparus aurata*) growth performance, feed utilisation, apparent digestibility of nutrients and nutrient outputs to the environment. Four isonitrogenous and isoenergetic diets were formulated: a control diet (CTRL) similar to a commercial feed, and three experimental diets containing, as main protein sources, plant by-products, glutens and concentrates (PLANT); processed animal proteins (PAP); or micro/macroalgae, insect meals and yeast (EMERG). Diets were tested in triplicate during 80 days. The EMERG treatment resulted in lower fish growth performance, higher FCR and lower nutrient and energy retentions than the other treatments. The lowest protein digestibility was found for the EMERG diet, which caused increased nitrogen losses. The PLANT and PAP treatments resulted in better fish growth performance, higher nutrient and energy retentions, and lower FCR than the CTRL treatment. The significant improvement in FCR found for fish fed PLANT and PAP diets and the high protein digestibility of these diets contribute towards minimizing the environmental impacts of seabream production.

## 1. INTRODUCTION

Aquaculture production is expected to continue to grow, as part of the solution to provide food for more than nine billion people in 2050 (FAO, 2018). However, growth must take into account the sustainability principles. Consumers are aware and concerned with the environmental impacts of the aquaculture industry, such as the use of limited and finite resources (such as fishmeal and fish oil) or the impact of nutrient discharges to the aquatic environment.

Research efforts in the last decades resulted in reduced feed conversion ratios in fish and in significant levels of fishmeal and fish oil replacement in aquafeeds (Olsen and Hasan, 2012). However, most of the research has focused on replacing fishmeal by terrestrial plant ingredients. At a short-medium term, this strategy places aquaculture in competition with biofuel and agricultural production for food (Olsen and Hasan, 2012), which is also expected to augment due to population increase. Furthermore, researchers and feed producers advert that a wider range of alternative ingredients to fishmeal are needed to fulfil the expected increase in aquaculture production (Matos, Dias, Dinis, & Silva, 2017).

Legislation constraints limit the use of alternative protein sources in aquafeeds in Europe. The use of terrestrial animal proteins in feed for all farmed animals was banned, in the European Union (EU) in 2001, to eradicate transmissible spongiform encephalopathies (European Parliament, 2001). Correctly categorised processed animal proteins (PAP) from non-ruminant origin (*e.g.*, porcine blood meal, poultry by-product meal and feather meal) were re-introduced in aquafeeds in EU in 2013 (European Comission, 2013). Due to the ban, research on the utilisation of PAP in aquafeeds was stagnated for several years within EU countries. These PAP are valuable feed ingredients, as they are waste from food production and are available in large amounts in the EU (Jędrejek, Levic, Wallace, & Oleszek, 2016). Recent studies indicate that these are suitable alternatives to fishmeal in practical fish diets (Campos, Matos, Marques, & Valente, 2017; Karapanagiotidis, Psofakis, Mente, Malandrakis, & Golomazou, 2019; Lu, Haga, & Satoh, 2015; Wang, Wang, Ji, Han, & Li, 2015; Wu, Ren, Chai, Li, & Wang, 2018).

In recent years, insects were considered as a new and renewable protein source for animal feed. Insects require fewer natural resources (such as land and water) than plants and can be reared on by-products, thus converting abundant low-cost organic waste into animal biomass that is rich in proteins and suitable for use in fish feed (Barroso *et al.*, 2014). In the EU, the use of insects as aquaculture feed ingredient was allowed in July 2017 (European Commission, 2017). Research results on the use of insects as partial replacement for fishmeal in fish feed are encouraging (Gasco *et al.*, 2016; Kroeckel *et al.*, 2012; Magalhães *et al.*, 2017; Sánchez-Muros *et al.*, 2016).

Other ingredients, such as micro- and macroalgae or one-cell organisms (for instance, yeasts), have started to emerge as promising alternative protein sources for fish feed and research on the use of these ingredients is increasing (de Cruz, Lubrano, & Gatlin, 2018; Øverland, Karlsson, Mydland, Romarheim, & Skrede, 2013; Vidakovic *et al.*, 2016, Vizcaíno, Mendes *et al.*, 2016a).

Novel formulations for fish feed need to evaluate their cost/benefit, an equation that have to balance raw material quality, cost and availability, but most importantly, the ramifications of integration of new ingredients for fish welfare, since the intestine is the primary target of dietary changes and challenges, thus emphasizing the fact that the integrity of the intestinal epithelium is fundamental to secure nutrient absorption and defense against pathogenic factors. Electrophysiological observations are a useful tool to evaluate alterations in active intestinal transport in fish and may be used as an indicator of changes in intestinal selectivity and integrity due to feeding with alternative diets (de Rodrigáñez, Fuentes, Moyano, & Ribeiro, 2013; Estensoro *et al.*, 2016; Vidakovic *et al.*, 2016).

Gilthead seabream (*Sparus aurata*) is one of the main farmed fish species in the Mediterranean area (FEAP, 2017). The future increase in aquaculture production must support the sustainable principles based on social, environmental and economical pillars. Therefore, this study aims to test alternative diet formulations for gilthead seabream production, with low levels of fishmeal inclusion and using non-conventional feed ingredients of animal and plant origin that are not used directly for human consumption. The objective of this study was to evaluate these alternative diets on gilthead seabream growth performance, feed utilisation, apparent digestibility of nutrients and nutrient outputs to the environment.

## 2. MATERIAL AND METHODS

### 2.1. Experimental diets

For this experiment, four isonitrogenous (crude protein: ~500 g/kg dry matter [DM]) and isoenergetic (gross energy: ~21.5 MJ/kg DM) diets were formulated with practical ingredients (Table 1). The control diet (CTRL) formulation was similar to a commercial gilthead seabream feed, using fishmeal and soy as main protein sources (350 g/kg of fishmeal, from which 50 g/kg consisted of fishmeal produced using fisheries by-products and 210 g/kg of soy products). The experimental diets were formulated to include a minimum amount of fair quality fishmeal produced from fish by-products (to ensure pellet palatability) and to maximise the use of alternative ingredients and/or by-products not directly used for human consumption. The main protein sources in the experimental diets were: plant-derived ingredients, including by-products, glutens and concentrates (PLANT diet); processed animal proteins, such as poultry meal, feather meal hydrolysate and porcine blood meal (PAP diet); and a mixture of emergent ingredients, such as micro- and macroalgae, insect meals and yeast (EMERG diet). Diets were supplemented whenever necessary with amino acids and inorganic phosphorus to fulfil the known nutritional requirements (indispensable amino acids and phosphorus) of juvenile gilthead seabream (*S. aurata*). The detailed formulation is presented in Table 1. Samples of each diet were collected and analysed for proximate composition (Table 1), amino acid (Table 2) and fatty acid (Table 3) content.

Diets (pellet size 2.0 mm) were manufactured at SPAROS Lda. (Olhão, Portugal) by extrusion by means of a pilot-scale twin-screw extruder (CLEXTRAL BC45, Clextral) with a screw diameter of 55.5 mm and temperature ranging from 105 to 110 ºC. Upon extrusion, all batches of extruded feeds were dried in a vibrating fluid bed dryer (model DR100, TGC Extrusion). Following drying, pellets were allowed to cool at room temperature and subsequently the oil fraction was added under vacuum coating conditions in a Pegasus vacuum mixer (PG-10VCLAB, DINNISEN). During the trial, all experimental diets were stored at room temperature, in a cool and aerated room.

Additionally, to measure apparent digestibility by the indirect method, 5 kg of each diet was reground, chromic oxide was incorporated at 10 g/kg and the mixtures were dry-pelleted (screen diameter: 4.5 mm), using a steamless pelleting machine (CPM-300).

## 2.2. Growth trial

All animal manipulations were carried out in compliance with the European (Directive 2010/63/EU) and Portuguese (Decreto-Lei nº 113/2013 de 7 de Agosto) legislation for the use of laboratory animals. All animal protocols were performed under Group-C licenses from the Direção-Geral de Alimentação e Veterinária, Ministério da Agricultura, Florestas e Desenvolvimento Rural, Portugal.

Gilthead seabream juveniles were obtained from Maresa – Mariscos de Esteros, SA. and transported to Ramalhete Experimental Research Station of CCMAR. Fish were adapted to the new conditions for several weeks, in a flow-through system with aeration, during which they were fed a commercial diet until the trial.

Twelve homogenous groups of 27 gilthead seabream juveniles with a mean initial body weight of 17.6 ± 1.4 g were stocked in circular plastic tanks (volume: 110 L; initial density: 5.3 kg/m$^3$), supplied with flow-through aerated seawater (temperature: 20.2 ± 1.4 °C; salinity: 36.2 ± 1.0 psu; dissolved oxygen in water above 80% of saturation) and subjected to natural photoperiod changes through autumn conditions (end September until beginning of December). Each diet was randomly assigned to triplicate tanks and tested over 80 days. Fish were hand fed to apparent satiety, two times a day (once on Sundays). Apparent feed intake was recorded and utmost care was taken to avoid feed losses. Mortality was monitored daily.

Anaesthetised (2-phenoxyethanol; 0.5 mL/L) fish were individually weighed at the beginning of the experiment and a group of 12 fish from the initial stock were euthanized by lethal anaesthesia (2-phenoxyethanol; 1.5 mL/L), pooled together and stored at -20 ºC until whole-body composition analysis. To monitor growth and feed utilisation, the fish were bulk weighed after 4 weeks under mild anaesthesia. At the end of the growth trial, each tank was bulk weighed and five fish from each tank were euthanized by lethal anaesthesia, individually weighed, pooled together and stored at -20 ºC for subsequent analysis of whole-body composition. Five additional fish from each tank were euthanized by lethal anaesthesia, individually weighed and liver and viscera were collected and weighed individually for analysis of somatic indexes. All samples were collected after 24 hr of fasting.

## 2.3. Apparent digestibility measurements

At the end of the growth trial and following all associated samplings, 12 groups of gilthead seabream (mean body weight: 63.6 ± 1.2 g; 15 fish per tank) were used to determine the apparent digestibility coefficients (ADC) of the dietary components by the indirect method, using 10 g/kg chromic oxide as a dietary inert tracer. Triplicate groups of fish were stocked in cylinder-conical tanks (volume: 100 L; temperature: 20.1 ± 0.1 °C; salinity: 35.4 ± 0.8 psu), in which the outlet water run through a recipient adapted to serve as faeces settling decantation system. Fish were fed with the experimental diets once a day. Fish were allowed to adapt to the new conditions during 5 days before faeces collection started. Once the experiment started, fish continued to be fed once a day during 30 min. After that period, tanks were thoroughly cleaned to remove any uneaten feed. Faecal samples were collected daily before feeding over 4 weeks. After daily collection, faeces were frozen at -20 °C.

Apparent digestibility coefficients of the dietary nutrients and energy were calculated as follows (Maynard, Loosli, Hintz, & Warner, 1979):

$$\text{ADC}\ (\%) = 100 \times \left[ 1 - \frac{\text{dietary } Cr_2O_3 \text{ level}}{\text{faecal } Cr_2O_3 \text{ level}} \times \frac{\text{faecal nutrient or energy level}}{\text{dietary nutrient or energy level}} \right]$$

ADC of dry matter was calculated as:

$$\text{ADC}\ (\%) = 100 \times \left[ 1 - \frac{\text{dietary } Cr_2O_3 \text{ level}}{\text{faecal } Cr_2O_3 \text{ level}} \right]$$

## 2.4. Short-circuit current (Isc) measurements

At the end of the growth trial and following all associated samplings, fish were kept on the same feeding regimes for one additional week and feed was withheld for 36 hr before sampling to ensure emptiness of the intestinal canal. On the day of the experiments, fish were anaesthetised with 2-phenoxyethanol (0.5 mL/L), sacrificed by decapitation, and the anterior intestine was collected, isolated and mounted in Ussing chambers as previously described (Carvalho, Gregório, Power, Canário, & Fuentes, 2012). Briefly, tissue was placed on a tissue holder of 0.25 $cm^2$ and positioned between two half-chambers containing 2 mL of serosal physiological saline (NaCl, 160 mM; $MgSO_4$, 1 mM; $NaH_2PO_4$, 2 mM; $CaCl_2$, 1.5 mM; $NaHCO_3$, 5 mM; KCl, 3 mM; glucose, 5.5 mM; HEPES (4-(2-Hydroxyethyl)piperazine-1-ethanesulfonic acid, N-(2-Hydroxyethyl)piperazine-N′-(2-ethanesulfonic acid)), 4mM), at a pH of 7.80. During the experiments, the tissue was bilaterally gassed with humidified air and the temperature was maintained at 20 °C. Short-circuit current (Isc, $\mu A/cm^2$) was monitored by clamping of epithelia to 0 mV. Epithelial resistance (Rt, $\Omega \cdot cm^2$) was manually calculated (Ohm's law) using the current deflections induced by a 2-mV pulse of 3 s every minute. Voltage clamping and current injections were performed by means of VCCMC2 amplifiers (Physiologic Instruments). Bioelectrical parameters for each tissue were recorded onto a computer by means of a Lab-Trax-4 acquisition system (World Precision Instruments) using LabScribe3 (iWorx Systems Inc.). For the purpose of presentation, data of individual fish were followed for 90 min after mounting and calculated values at 10-min intervals were averaged for each individual tissue.

## 2.5. Analytical methods

Whole-body fish samples from each tank were pooled together ($n$ = 3 per treatment) and all samples (whole-body, diets and faeces) were ground before analysis. Chemical analysis followed standard procedures of the Association of Official Analytical Chemists (AOAC, 2006) and were run in duplicates. Moisture (105 °C for 24 hr) and ash (combustion at 550 °C for 12 hr) content were determined in samples of diets and whole-body fish. Freeze-dried samples from diets, faeces and whole-body fish were analysed for crude protein (N x 6.25) content using an elemental analyser (Elementar Vario EL III), gross energy by combustion in an adiabatic bomb calorimeter (Werke C2000, IKA), and total phosphorus content by digestion at 230 °C in a Kjeldatherm block digestion unit followed by digestion at 75 °C in a water bath and absorbance determination at 820 nm (adapted from AFNOR V 04-406). Diet and faeces samples were analysed for crude fat content by petroleum ether extraction using a Soxtherm Multistat/SX PC (Gerhardt) and chromic oxide content was determined according to Bolin, King, and Klosterman (1952), after perchloric acid digestion.

Freeze-dried samples from experimental diets and faeces were hydrolysed (6 M HCl at 116 ºC over 48 hr in nitrogen-flushed glass vials) before total amino acid analyses. All samples were then pre-column derivatised with Waters AccQ Fluor Reagent (6-aminoquinolyl-N-hydroxysuccinimidyl carbamate) using the AccQ Tag method (Waters). Analyses were done by ultra-high-performance liquid chromatography (UPLC) in a Waters reversed-phase amino acid analysis system, using norvaline as an internal standard. The resultant peaks were analysed with EMPOWER software (Waters).

Fatty acids methyl esters (FAME) of diets were obtained by direct acid-catalysed methylation (Lepage & Roy, 1984), using nonadecanoic acid (19:0, Matreya LLC) as internal standard. FAME were analysed by gas chromatography using a Shimadzu GC-2010 (Shimadzu Europe GmbH) chromatograph, equipped with a flame-ionization detector and an Omegawax 250 capillary column (30 m x 0.25 mm i.d. x 0.25 µm film thickness; Supelco). The initial oven temperature of 150 ºC was held for 7 min, increased to 170 ºC at 3 ºC/min and held for 25 min, and then increased to 220 ºC at 3 ºC/min and held for 30 min. The temperatures of injector and detector were maintained at 250 and 260 ºC, respectively. Helium was used as carrier gas. The injected volume was 1.0 µL and the split ratio 1:100. FAME identification was achieved by comparison with the retention times of standard mixtures (Supelco 37 Component FAME Mix, Bacterial Acid Methyl Ester Mix, PUFA No. 1, PUFA No. 2, and PUFA No. 3, Sigma-Aldrich Co. LLC and GLC-110 Mixture; Matreya LLC).

## 2.6. Calculations

Growth indexes and nutrient retention parameters were calculated as follows:

Daily growth index (DGI): 100 × ($FBW^{1/3}$ – $IBW^{1/3}$) x $days^{-1}$, where FBW and IBW are the final and initial body weights, respectively.

Feed conversion ratio (FCR): apparent feed intake x wet weight $gain^{-1}$, where wet weight gain is: FBW – IBW.

Average body weight (ABW): (IBW + FBW) / 2.

Daily voluntary feed intake (VFI, %·$day^{-1}$): 100 x apparent feed intake x $ABW^{-1}$ x $days^{-1}$.

Protein efficiency ratio (PER): wet weight gain × crude protein $intake^{-1}$.

Hepatosomatic index (HSI, %): 100 × liver weight × $FBW^{-1}$.

Viscerosomatic index (VSI, %): 100 × viscera weight × $FBW^{-1}$.

Nutrient retention (% intake): 100 x (FBW x final whole-body nutrient content – IBW x initial whole-body nutrient content) / (nutrient intake).

Nitrogen (N) or phosphorus (P) gain (mg N or P·kg$^{-1}$ fish·day$^{-1}$): (final whole-body N or P content – initial whole-body N or P content) x ABW$^{-1}$ x days$^{1}$.

Faecal N or P losses (mg N or P·kg$^{-1}$ fish·day$^{-1}$): [N or P intake x (100 - N or P ADC %)] x ABW$^{-1}$ x days$^{-1}$.

Metabolic N or P losses (mg N or P·kg$^{-1}$ fish·day$^{-1}$): N or P intake - (N or P gain + faecal N or P losses).

### 2.7. Statistical analysis

Data are expressed as means ± standard deviation. Results expressed as percentages were transformed (arcsine square root) before statistical analysis, according to Ennos (2007). Data were checked for normal distribution and homogeneity of variances before analysis. All data were subjected to one-way analysis of variance (ANOVA). When significant differences were detected, the Tukey's multiple-comparison test was used to assess differences among groups. Differences were considered significant when $p < .05$. All statistical tests were performed using RStudio software Version 1.1.453.

## 3. RESULTS

During the 80 days of experiment, survival was high and only one fish from the CTRL treatment and one from the EMERG treatment died.

Fish from all treatments more than doubled their initial weight during the experiment (Table 4). Gilthead seabream fed the EMERG diet presented the lowest growth performance, reflected in significantly lower values for final body weight, DGI and PER than those obtained in the other treatments. The highest FCR (2.0) was also found for the EMERG treatment, which was significantly different from the other treatments. Fish fed PLANT and PAP diets presented higher PER and lower FCR than fish fed CTRL diet. Feed intake (VFI) was significantly higher in fish fed the EMERG than the PLANT and PAP diets. The hepatosomatic index was only significantly higher in fish fed the EMERG diet (1.9 %) compared with fish fed the CTRL diet (1.5 %), while the viscerosomatic index was significantly higher in the EMERG treatment than in the other treatments.

Whole-body composition at the end of the experiment was not significantly affected by the dietary treatments (Table 5). Average values for moisture ranged from 643 to 660 g/kg and for protein ranged from 167 to 174 g/kg (wet weight). Phosphorus and energy contents were also similar among treatments.

Protein retention (Table 6) in gilthead seabream juveniles fed PLANT and PAP diets was significantly higher (~28%) than in fish fed CTRL (~23%) and EMERG (~20%) diets. Phosphorus retention was not significantly affected by the dietary treatment. Energy retention were significantly different in all treatments, being higher in fish fed PAP and PLANT diets than in fish fed the CTRL diet. Fish fed the EMERG diet presented the lowest energy retention.

Concerning diet digestibility (Table 7), ADC of dry matter, lipids and energy were similar for CTRL, PLANT and PAP diets, but significantly lower for the EMERG diet. Protein digestibility was also significantly lower for the EMERG diet (~70%) when compared with the other diets (>90%). Phosphorus digestibility was significantly higher for the PLANT diet than for the CTRL and PAP diets (Table 7). Apparent digestibility coefficients for amino acids were generally lower in the EMERG diet than in the other diets (Table 8), with values usually around 75-85%. The ADC of amino acids were usually significantly higher for the PAP and CTRL diets than for the PLANT diet (Table 8).

Whole-body composition analysis combined with data on ADC of diets allowed the calculation of daily nitrogen (Figure 1) and phosphorus (Figure 2) balances. Daily nitrogen gain (325 ± 2 to 391 ± 60 mg N $kg^{-1}$ fish·$day^{-1}$) was not significantly different among treatments (Figure 1). Fish fed the CTRL diet had significantly higher metabolic nitrogen losses (1117 ± 10 mg N·$kg^{-1}$ fish·$day^{-1}$) than those fed the other diets (858 ± 11 to 916 ± 38 mg N·$kg^{-1}$ fish·$day^{-1}$). Faecal nitrogen losses were significantly different in all treatments (ranging from 86 ± 2 to 512 ± 4 mg N·$kg^{-1}$ fish·$day^{-1}$), with the highest value found for the EMERG treatment and the lowest for the PLANT treatment.

Daily phosphorus gain (90 ± 10 to 95 ± 12 mg P·$kg^{-1}$ fish·$day^{-1}$) was not affected by the dietary treatments (Figure 2). Metabolic phosphorus losses were significantly higher in fish fed the EMERG diet (70 ± 13 mg P·$kg^{-1}$ fish·$day^{-1}$) compared with fish fed the CTRL (40 ± 9 mg P·$kg^{-1}$ fish·$day^{-1}$) and PAP (29 ± 12 mg P·$kg^{-1}$ fish·$day^{-1}$) diets. Faecal phosphorus losses were significantly higher in the fish fed the CTRL (94 ± 2 mg P·$kg^{-1}$ fish·$day^{-1}$) than the PLANT diet (64 ± 1 mg P·$kg^{-1}$ fish·$day^{-1}$), with intermediate values for the other treatments.

The values of transepithelial resistance decreased in fish fed the alternative diets compared with fish fed the CTRL diet (Figure 3). However, significant differences were only found between the CTRL (188 ± 53 Ω·$cm^2$) and the EMERG (109 ± 35 Ω·$cm^2$) treatments. The values of short-circuit current (Isc) measured in the intestine of gilthead seabream were all negative (Figure 3), showing that current direction did not change by feeding alternative diets. No significant differences in Isc were found among diets. Nevertheless, Isc values in fish fed the CTRL diet (-4.2 ± 2.5 µA/$cm^2$) were in average lower than the values found for the other treatments (-6.2 ± 6.0, -8.9 ± 4.1 and -8.2 ± 5.5 µA/$cm^2$, respectively for PLANT, PAP and EMERG treatments) and the dispersion of values was also much higher in fish fed the alternative diets compared with the CTRL diet.

## 4. DISCUSSION

Fishmeal replacement in gilthead seabream diets is not a novel concept and previous studies showed the potential of using a mixture of plant proteins to replace high dietary levels of fishmeal, given that a dietary balanced amino acid profile and minimal inclusion of anti-nutritional factors were guaranteed (*e.g.*, Dias *et al.*, 2009, Monge-Ortiz *et al.*, 2016). The CTRL diet used in this study had a similar formulation to a commercial diet used nowadays in gilthead seabream production, which already includes a large amount of plant ingredients as protein sources. However, the concept beyond the scope of this study was to minimise the use of ingredients that may be used directly for human consumption and focuses more on non-conventional feed ingredients, from both animal and plant origin. Therefore, the PLANT diet formulation included several plant glutens and concentrates, and also plant by-products, such as wheat germ, guar meal, carob germ and wheat DDGS. Studies on the inclusion of these plant by-products in aquafeeds are relatively recent and scarce for carnivorous fish species, but in general report good perspectives for their use (*e.g.*, Campos, Matos, Aragão, Pintado, & Valente, 2018; Couto *et al.*, 2016, Diógenes, Castro, Miranda, Oliva-Teles, & Peres, 2018; Pach and Nagel, 2018). To the best of our knowledge, there are no available studies reporting the mixture of so many plant by-products in fish diets. Studies testing increasing levels of single by-products in gilthead seabream diets have shown that, for instance, carob seed germ meal can be included at a maximum level of 340 g/kg (Martínez-Llorens, Baeza-Ariño, Nogales-Mérida, Jover-Cerdá, & Tomás-Vidal, 2012) and corn DDGS can be included at 350 g/kg (Diógenes *et al.*, 2019) without negative effects on growth or protein and energy retentions. The current study showed that fish fed the PLANT diet performed better than those fed the CTRL diet, which included relatively high levels of fishmeal. Therefore, the mixture of several plant by-products seems an efficient strategy to formulate more sustainable diets for gilthead seabream that ultimately result in higher protein and energy retentions.

Previous studies showed that gilthead seabream was able to cope with dietary fishmeal replacement by rendered animal proteins. Recent studies in gilthead seabream recommend the dietary replacement of up to 50% of fishmeal by poultry meal (Davies *et al.*, 2019, Karapanagiotidis *et al.*, 2019) and inclusion of feather meal at 50 to 100 g/kg and blood meal at 100 g/kg (Davies *et al.*, 2019; Nogueira, Cordeiro, Andrade, & Aires, 2012). Protein digestibility was lower in diets containing feather meal than in diets containing poultry meal or spray-dried haemoglobin meal for several carnivorous fish species, including gilthead seabream (Davies, Gouveia, Laporte, Woodgate, & Nates, 2009). The results of the current study showed that although the protein digestibility of the PAP diet was slight lower compared with the CTRL diet, gilthead seabream performance was slightly better when fed the PAP than the CTRL diet. Furthermore, contrary to previous studies (Karapanagiotidis *et al.*, 2019), protein retention was better in fish fed the PAP than the CTRL diet. As suggested by Davies *et al.* (2019), the blending of different PAP proved to be a successfully strategy to reduce nutrient imbalances and prevent problems related with diet palatability, allowing a high level of fishmeal replacement in seabream diets. The mixing of poultry meal, feather meal and blood meal has also proven successfully as a strategy to reduce fishmeal inclusion in rainbow trout (*Oncorhynchus mykiss*) diets, a carnivorous fish species like gilthead seabream (Gaylord, Sealey, Barrows, Myrick, & Fornshell, 2017).

Contrarily to the other experimental diets, fish performance was worse when fed the EMERG diet. These results are probably related with the lower digestibility of the EMERG diet compared with the CTRL and the other experimental diets, indicating that a significant part of nutrients and energy was not being utilised and retained by the fish due to faecal losses.

At this point, the overall lower fish performance and nutrient retention and lower diet digestibility found in the EMERG treatment cannot be attributed to a single cause. The EMERG diet included a mixture of different emergent ingredients in fish diet formulation, as a strategy to prevent possible dietary deficiencies. However, available data on these novel ingredients are still scarce and results can be contradictory. For instance, for brewer's yeast, dietary inclusion levels up to 550 g/kg did not affect growth performance and even increased protein retention of European seabass (*Dicentrarchus labrax*; Oliva-Teles & Gonçalves, 2001), while dietary inclusion levels of 400 g/kg negatively affected growth performance of Atlantic salmon (*Salmo salar*; Øverland *et al.*, 2013). Furthermore, although no negative effects on growth performance of European seabass were perceptible, nutrient digestibility was affected at dietary inclusion levels equal or above 219 g/kg (Oliva-Teles & Gonçalves, 2001). Nutrient utilisation and/or digestibility were also significantly lower in Atlantic salmon and Artic charr (*Salvelinus alpinus*) fed diets containing yeast at 400 and 298 g/kg, respectively (Øverland *et al.*, 2013, Vidakovic *et al.*, 2016). In our study, the inclusion level of brewer's yeast in the EMERG diet was low (50 g/kg), so it seems unlikely that this ingredient affected gilthead seabream growth performance. However, it remains unclear if nutrient digestibility could have been affected.

Concerning insect meal, its inclusion in fish diets is relatively recent and has led to different results. For instance, growth and feeding efficiency were not affected by inclusion of *Tenebrio molitor* meal at 250 g/kg in diets for European seabass (Gasco *et al.*, 2016), nor by inclusion at 500 g/kg in diets for rainbow trout (Belforti *et al.*, 2015). However, growth was significantly lower when 210 g/kg of *Tenebrio molitor* meal was included in diets for Nile tilapia (*Oreochromis niloticus*; Sánchez-Muros *et al.*, 2016). Negative effects of dietary insect meal inclusion on nutrient digestibility, especially on protein digestibility, have been reported, which seems to be related with chitin levels (Belforti *et al.*, 2015; Kroeckel *et al.*, 2012,). Other studies reported no adverse effects of dietary insect meal inclusion on protein digestibility (Gasco *et al.*, 2016, Magalhães *et al.*, 2017). Therefore, the chitin levels in the EMERG diet may have affected the apparent digestibility of nutrients by gilthead seabream. Nevertheless, the insect meals used in the current study present low chitin content (Barroso *et al.*, 2014), and dietary inclusion levels were much lower than those reported to affect negatively fish performance (Kroeckel *et al.*, 2012, Sánchez-Muros *et al.*, 2016).

As for the other ingredients included in the EMERG diet, macroalgae inclusion at levels of 50 g/kg, similar to the ones used in the current study, did not affect growth and FCR in gilthead seabream juveniles, but reduced the amount of intestinal proteases (Vizcaíno, Mendes, *et al.*, 2016a). Concerning microalgae, dietary inclusion of *Scenedesmus* does not seem to cause any adverse effect on gilthead seabream (Vizcaíno *et al.*, 2014), but other microalgae (*Tetraselmis suecica* and *Tisochrysis lutea*) resulted in negative effects (Vizcaíno *et al.*, 2016b). Data on the digestibility of each individual microalgae species is still clearly needed. Contrarily to previous studies (de Cruz *et al.*, 2018; Vizcaíno, Saéz, *et al.*, 2016b), the current study used a diet containing a mixture of different microalgae at higher percentage levels, while fishmeal inclusion levels were kept to a minimum (50 g/kg). This may be a possible explanation for the reduced growth and nutrient digestibility found in fish fed the EMERG diet. Knowledge generated in the current study illustrates some of the challenges that still need to be overcome in order to potentiate the use of microalgae as valuable ingredients in the formulation of cost-effective feeds for fish.

Whole-body composition was not apparently affected by the dietary treatments, although due to analytical problems whole-body fish content could not be analysed. Previous studies showed that fat content was higher in gilthead seabream fed diets containing blood meal and feather meal (Nogueira *et al.*, 2012), while it was not significantly affected when fish were fed diets containing feather meal, spray-dried haemoglobin or poultry meal alone (Davies *et al.*, 2019, Karapanagiotidis *et al.*, 2019). Additionally, hepatosomatic and viscerosomatic indexes have been reported to be significantly higher in fish fed insect meal-based diets (Gasco *et al.*, 2016, Sánchez-Muros *et al.*, 2016). The same was observed in the current experiment when fish were fed the EMERG diet, indicating an increase in visceral fat deposition. Further experiments are necessary to clarify this issue, since an increase in fish body fat may affect negatively consumers' acceptability.

Alternative formulations are nowadays being directed to an eco-friendly approach (Matos *et al.*, 2017) that targets a reduction of nutrient outputs to the environment. Nitrogen discharges to the environment result from both faecal losses, mostly related with low dietary protein digestibility, but also from metabolic losses, resulting, from instance, from an imbalanced dietary amino acid profile. The nitrogen balance calculated in the current study indicates that nitrogen discharges were mostly related to the low protein digestibility found for the EMERG diet, since metabolic losses were lower in fish fed the alternative diets compared with the CTRL. Therefore, the lower protein digestibility in the EMERG diet had not only a detrimental effect on fish growth performance, but also a negative environmental impact. Furthermore, results from phosphorus balance show that phosphorus intake and outputs to the environment (sum of faecal and metabolic losses) were also higher in gilthead seabream fed the EMERG diet. Since phosphorus digestibility was high for this diet and fish phosphorus gain was similar in all the treatments, this suggests that supplementation with mono-calcium phosphate in the EMERG diet was probably not necessary. Clearly, there is room for future research to reduce the environmental impact of diets with alternative emergent ingredients.

On a positive way, feeding gilthead seabream with PLANT and PAP diets did not increase phosphorus outputs to the environment. In line with our results, soluble phosphorus excretion in gilthead seabream was significantly reduced when feeding plant protein-rich diets, resulting in lower phosphorus losses to the environment (Dias *et al.*, 2009). Likewise, feeding gilthead seabream with the PAP diet (containing poultry meal, blood meal and feather meal) did not increase phosphorus losses to the environment. Phosphorus waste outputs in Japanese seabass (*Lateolabrax japonicus*) increased with high inclusion levels of dietary poultry meal (Wang *et al.*, 2015), but similarly to our study, feeding giant croaker (*Nibea japonica*) on a diet based on poultry meal and feather meal did not increase phosphorus losses (Wu *et al.*, 2018).

The electrophysiological observations showed that feeding gilthead seabream with alternative diets did not evoke major modifications in active intestinal transport. The negative Isc found in the current study is associated with absorptive function, reflecting a proper functionality of the epithelium and the absence of inflammation/enteritis, usually associated with current inversions driven by diarrheic processes (de Rodrigáñez *et al.*, 2013). Nevertheless, the high Isc values found in gilthead seabream fed alternative diets suggest a higher potential cost for transcellular transport than for fish fed the CTRL diet. The results were not significantly different due to the high variation found among individuals fed the alternative diets, which indicates different individual physiological responses and suggests different abilities to cope with the alternative dietary ingredients. Furthermore, the electrophysiological results showed that feeding gilthead seabream with alternative diets decreased transepithelial resistance, an indicator of intestinal selectivity/integrity, which has been previously observed in gilthead seabream fed diets with high levels of fishmeal replacement by plant proteins (Estensoro *et al.*, 2016). Although fishmeal inclusion in alternative diets was only 50 g/kg, Rt values were only significantly different between the CTRL and the EMERG diets, with PLANT and PAP diets presenting intermediate values. This result corroborates the potential use of by-products from terrestrial plant and animal origin to replace dietary fishmeal. Tissue resistance is considered the electrical expression of barrier function in epithelial systems (de Rodrigáñez *et al.*, 2013) and the decrease in Rt values indicate changes in the paracellular pathway integrity, which may result in a favourable route for the access of pathogens (Zhang, Hornef, & Dupont, 2015). The negative effects were clearly more pronounced with the use of alternative emergent ingredients (EMERG diet). Previous studies reported an impaired intestinal barrier function due to increased paracellular permeability and decreased transepithelial resistance in Artic charr fed yeast-based diets (Vidakovic *et al.*, 2016). Therefore, it is possible that yeast inclusion in the EMERG diet affected the fish intestinal barrier function but cannot rule out the possible effect of the other novel ingredients used in the formulation. Further research is necessary to evaluate and prevent the impacts of these novel emergent ingredients on fish health and welfare.

In summary, algae and insect meals seem a potential sustainable option to replace fishmeal and terrestrial plant proteins, but more detailed studies are still necessary to optimise inclusion levels. This is of utmost importance to avoid physiological impairments that result in the loss of intestinal integrity and lower feed utilisation, which translates in lower growth performance. The significant improvement in feed conversion ratios of fish fed PLANT and PAP diets contribute towards minimising the environmental impacts of the aquaculture industry and pave the way to formulate diets using resources that are not directly used for human consumption.

## ACKNOWLEDGEMENTS

This work was supported by the Portuguese Foundation for Science and Technology (Ministry of Science and Higher Education, Portugal and European Social Funds) through projects UID/Multi/04326/2019 to CCMAR, PTDC/MAR-BIO/3034/2014 to JF, and contract DL57/2016/CP1361/CT0033 to CA.

## Data Availability Statement

Research data are not shared.

**TABLES**

**TABLE 1** Formulation and proximate composition of the experimental diets.

| *Ingredients (g/kg)* | CTRL | PLANT | PAP | EMERG |
|---|---|---|---|---|
| Fishmeal LT Diamante [1] | 300 | 0 | 0 | 0 |
| White fishmeal (by-products) [2] | 50 | 50 | 50 | 50 |
| Soybean meal [3] | 100 | 0 | 0 | 0 |
| Soy protein concentrate [4] | 110 | 0 | 0 | 0 |
| Corn gluten meal [5] | 100 | 140 | 100 | 0 |
| Rapeseed meal [6] | 50 | 50 | 50 | 0 |
| Wheat gluten [7] | 30 | 100 | 0 | 45 |
| Wheat meal [8] | 70 | 0 | 155 | 0 |
| Guar meal [9] | 0 | 140 | 0 | 50 |
| Pea protein concentrate [10] | 0 | 100 | 0 | 0 |
| Carob germ [11] | 0 | 69 | 0 | 0 |
| Wheat germ [12] | 0 | 50 | 0 | 57 |
| Wheat DDGS [13] | 0 | 46 | 0 | 0 |
| Poultry meal [14] | 0 | 0 | 390 | 0 |
| Feather meal hydrolysate [15] | 0 | 0 | 40 | 0 |
| Porcine blood meal [16] | 0 | 0 | 37 | 0 |
| *Chlorella* [17] | 0 | 0 | 0 | 150 |
| *Scenedesmus* [18] | 0 | 0 | 0 | 150 |
| *Spirullina* [19] | 0 | 0 | 0 | 70 |
| Tenebrio meal [20] | 0 | 0 | 0 | 100 |
| Locust meal [21] | 0 | 0 | 0 | 50 |
| Macroalgae mix [22] | 0 | 0 | 0 | 50 |
| Brewer's yeast [23] | 0 | 0 | 0 | 50 |
| Pea starch [24] | 30 | 30 | 30 | 0 |
| Fish oil [25] | 133 | 140 | 102 | 113 |
| Vitamin & mineral premix [26] | 10 | 10 | 10 | 10 |
| Soy lecithin [27] | 10 | 10 | 10 | 10 |
| Guar gum [28] | 2 | 2 | 2 | 2 |
| Antioxidant powder [29] | 2 | 2 | 2 | 2 |
| Sodium propionate [30] | 1 | 1 | 1 | 1 |
| Mono-calcium phosphate [31] | 0 | 23 | 5 | 18 |
| DL-Methionine [32] | 2 | 9 | 3 | 8 |
| Lysine-sulphate [33] | 0 | 15 | 10 | 8 |
| L-Threonine [34] | 0 | 7 | 2 | 2 |
| L-Histidine [35] | 0 | 4 | 1 | 4 |
| L-Tryptophan [36] | 0 | 2 | 0 | 0 |
| | | | | |
| *Proximate composition (g/kg DM)* | | | | |
| Dry matter (DM) | 960 | 922 | 947 | 939 |
| Ash | 82 | 57 | 74 | 104 |
| Crude protein | 499 | 496 | 499 | 499 |
| Crude fat | 200 | 181 | 173 | 173 |
| Total phosphorus | 11 | 12 | 12 | 12 |
| Gross energy *(MJ/kg DM)* | 21.7 | 22.4 | 21.3 | 21.2 |

Note: All values are reported as mean of duplicate analysis.

[1] Fish meal NORVIK 70: 703 g/kg crude protein (CP), 58 g/kg crude fat (CF), Sopropêche.

[2] White fishmeal (by-products): 630 g/kg CP, 90 g/kg CF, Savinor UTS.

[3] Dehulled solvent extracted soybean meal: 470 g/kg CP, 26 g/kg CF, CARGILL.

[4] Soycomil P: 630 g/kg CP, 8 g/kg CF, ADM.

[5] Corn gluten meal: 610 g/kg CP, 60 g/kg CF, COPAM.

[6] Defatted rapeseed meal: 340 g/kg CP, 20 g/kg CF, Premix Lda.

[7] VITAL: 800 g/kg CP, 75 g/kg CF, Roquette Frères.

[8] Wheat meal: 102 g/kg CP, 12 g/kg CF, Casa Lanchinha.

[9] Korfeed 60: 580 g/kg CP, 65 g/kg CF, Sopropêche.

[10] Pea protein concentrate: 780 g/kg CP, 9 g/kg$^{-1}$ CF, Roquette Frères.

[11] Carob germ: 450 g/kg CP, 40 g/kg CF, Premix Lda.

[12] Wheat germ: 281 g/kg CP, 118 g/kg CF, GERMEN SA.

[13] Pannonia GOLD DDGS: 320 g/kg CP, 93 g/kg CF, Pannonia Ethanol Zrt.

[14] Poultry meal 65: 670 g/kg CP, 120 g/kg CF, Sonac.

[15] Feather meal hydrolysate: 842 g/kg CP, 104 g/kg CF, Savinor UTS.

[16] Porcine blood meal: 890 g/kg CP, 80 g/kg CF, Sonac.

[17] *Chlorella* sp.: 620 g/kg CP, 90 g/kg CF, Allmicroalgae.

[18] *Scenedesmus obliquus*: 480 g/kg CP, 82 g/kg CF, Allmicroalgae.

[19] *Spirullina*: 720 g/kg CP, 10 g/kg CF, Willows Ingredients.

[20] Tenebrio meal: 670 g/kg CP, 160 g/kg CF, Entomo Farm.

[21] Locust meal: 630 g/kg CP, 150 g/kg CF, ACRS.

[22] Macroalgae mix: 110 g/kg CP, 6 g/kg CF, Ocean Harvest.

[23] Brewer's yeast: 390 g/kg CP, 45 g/kg CF, PREMIX Lda.

[24] NASTAR: 3 g/kg CP, 1 g/kg CF, COSUCRA.

[25] Fish oil: Savinor UTS.

[26] PREMIX Lda: Vitamins (IU or mg/kg diet): DL-alpha tocopherol acetate, 100 mg; sodium menadione bisulphate, 25 mg; retinyl acetate, 20000 IU; DL-cholecalciferol, 2000 IU; thiamin, 30 mg; riboflavin, 30 mg; pyridoxine, 20 mg; cyanocobalamin, 0.1 mg; nicotinic acid, 200 mg; folic acid, 15 mg; ascorbic acid, 1000 mg; inositol, 500 mg; biotin, 3 mg; calcium panthotenate, 100 mg; choline chloride, 1000 mg; betaine, 500 mg. Minerals (g or mg/kg diet): cobalt carbonate, 0.65 mg; copper sulphate, 9 mg; ferric sulphate, 6 mg; potassium iodide, 0.5 mg; manganese oxide, 9.6 mg; sodium selenite, 0.01 mg; zinc sulphate,7.5 mg; sodium chloride, 400 mg; calcium carbonate, 1.86 g; excipient wheat middlings.

[27] Lecico P700IPM, LECICO GmbH.

[28] Guar gum: Seah International.

[29] Paramega PX, KEMIN EUROPE NV.

[30] Sodium propionate: Disproquímica.

[31] MCP: 220 g/kg P, 180 g/kg Ca, Fosfitalia.

[32] DL-Methionine: 990 g/kg, EVONIK Nutrition & Care GmbH.

[33] Biolys: L-lysine sulphate, 546 g/kg Lysine, EVONIK Nutrition & Care GmbH.

[34] L-Threonine: 980 g/kg, EVONIK Nutrition & Care GmbH.

[35] L-Histidine: 980 g/kg, AJINOMOTO.

[36] L-Tryptophan: 980 g/kg, EVONIK Nutrition & Care GmbH.

**TABLE 2** Amino acid composition of experimental diets.

| Amino acids | DIETS | | | |
|---|---|---|---|---|
| (g/kg DM) | CTRL | PLANT | PAP | EMERG |
| Arg | 32 | 36 | 32 | 32 |
| His | 10 | 13 | 11 | 14 |
| Lys | 28 | 26 | 30 | 29 |
| Thr | 18 | 18 | 18 | 19 |
| Ile | 19 | 15 | 16 | 17 |
| Leu | 39 | 32 | 36 | 29 |
| Val | 22 | 18 | 23 | 25 |
| Met | 13 | 14 | 12 | 15 |
| Phe | 22 | 20 | 20 | 19 |
| Cys | 2 | 2 | 2 | 2 |
| Tyr | 19 | 17 | 16 | 22 |
| Asx | 41 | 30 | 32 | 35 |
| Glx | 51 | 83 | 59 | 55 |
| Ala | 26 | 19 | 26 | 30 |
| Gly | 29 | 20 | 34 | 29 |
| Pro | 27 | 27 | 30 | 25 |
| Ser | 22 | 19 | 21 | 20 |
| Tau | 3 | 1 | 3 | 1 |

Note: All values are reported as mean of duplicate analysis.

**TABLE 3** Fatty acid composition of experimental diets.

| Fatty acids | DIETS | | | |
|---|---|---|---|---|
| (% of total) | CTRL | PLANT | PAP | EMERG |
| 14:0 | 3.9 | 3.4 | 3.0 | 3.4 |
| 16:0 | 18.0 | 18.0 | 19.8 | 19.6 |
| 17:0 | 0.8 | 0.7 | 0.6 | 0.7 |
| 18:0 | 4.3 | 4.2 | 5.5 | 4.0 |
| ∑ SFA | 29.4 | 28.6 | 31.1 | 30.3 |
| 16:1n-7 | 0.3 | 0.3 | 0.3 | 0.9 |
| 18:1n-9 | 15.2 | 15.2 | 19.8 | 17.0 |
| 18:1n-7 | 2.6 | 2.3 | 2.4 | 2.2 |
| 20:1n-9 | 2.7 | 1.7 | 1.6 | 1.6 |
| 22:1n-9 | 0.4 | 0.3 | 0.3 | 0.3 |
| ∑ MUFA | 30.1 | 28.1 | 31.4 | 27.9 |
| 18:2n-6 | 6.6 | 14.8 | 12.2 | 11.8 |
| 18:3n-3 | 1.2 | 1.6 | 1.2 | 3.7 |
| 18:4n-3 | 1.3 | 1.0 | 0.9 | 1.0 |
| 20:4n-6 | 1.2 | 1.1 | 1.4 | 0.9 |
| 20:5n-3 | 7.2 | 5.9 | 5.0 | 5.1 |
| 22:5n-3 | 1.2 | 0.9 | 0.9 | 0.8 |
| 22:6n-3 | 15.0 | 11.8 | 10.4 | 10.2 |
| ∑ PUFA | 36.3 | 39.4 | 34.2 | 36.1 |
| ∑ n-3 | 26.7 | 22.0 | 19.0 | 21.5 |
| ∑ n-6 | 8.4 | 16.3 | 14.2 | 13.5 |
| n-3/n-6 | 3.2 | 1.4 | 1.3 | 1.6 |

Note: All values are reported as mean of duplicate analysis.

Abbreviations: MUFA, monounsaturated fatty acids; PUFA, polyunsaturated fatty acids; SFA, saturated fatty acids.

**TABLE 4** Growth performance and somatic indexes of gilthead seabream juveniles fed the experimental diets for 80 days (IBW: 17.6 ± 1.4 g).

| | CTRL | PLANT | PAP | EMERG |
|---|---|---|---|---|
| FBW (g) | $59.88 \pm 8.03^{a}$ | $64.77 \pm 9.07^{a}$ | $62.94 \pm 8.94^{a}$ | $48.25 \pm 6.07^{b}$ |
| DGI | $1.64 \pm 0.05^{b}$ | $1.77 \pm 0.05^{a}$ | $1.72 \pm 0.01^{ab}$ | $1.30 \pm 0.05^{c}$ |
| FCR | $1.5 \pm 0.06^{b}$ | $1.3 \pm 0.01^{c}$ | $1.3 \pm 0.04^{c}$ | $2.0 \pm 0.08^{a}$ |
| VFI ($\%\cdot day^{-1}$) | $3.81 \pm 0.14^{bc}$ | $3.35 \pm 0.01^{a}$ | $3.32 \pm 0.14^{ab}$ | $4.89 \pm 0.20^{c}$ |
| PER | $1.37 \pm 0.05^{b}$ | $1.63 \pm 0.01^{a}$ | $1.60 \pm 0.05^{a}$ | $1.09 \pm 0.04^{c}$ |
| HSI (%) | $1.54 \pm 0.41^{b}$ | $1.64 \pm 0.25^{ab}$ | $1.78 \pm 0.36^{ab}$ | $1.89 \pm 0.29^{a}$ |
| VSI (%) | $4.54 \pm 0.54^{b}$ | $4.57 \pm 0.52^{b}$ | $4.74 \pm 0.70^{b}$ | $5.78 \pm 0.55^{a}$ |

Note: Values are means ± standard deviation ($n = 3$, except for FBW where $n = 30$ and HSI/VSI where $n = 15$). Different superscripts within the same row indicate significant differences ($p < .05$) among diets.

Abbreviations: DGI, daily growth index; FBW, final body weight; FCR, feed conversion ratio; HSI, hepatosomatic index; IBW, initial body weight; PER, protein efficiency ratio; VFI, daily voluntary feed intake; VSI, viscerosomatic index.

**TABLE 5** Whole-body composition (g/kg wet weight) of gilthead seabream juveniles fed the experimental diets for 80 days.

| Body composition | CTRL | PLANT | PAP | EMERG |
|---|---|---|---|---|
| Moisture | 660 ± 3 | 651 ± 9 | 643 ± 6 | 645 ± 16 |
| Ash | 35 ± 5 | 33 ± 3 | 34 ± 7 | 37 ± 5 |
| Protein | 167 ± 10 | 171 ± 14 | 173 ± 4 | 174 ± 4 |
| Phosphorus | 2.1 ± 0.3 | 1.8 ± 0.1 | 1.7 ± 0.2 | 2.3 ± 0.3 |
| Energy (MJ/kg) | 79 ± 1 | 81 ± 2 | 85 ± 1 | 82 ± 5 |

Note: Values are means ± standard deviation ($n$ = 3). Absence of superscripts within the same row indicates no significant differences ($p > .05$) among diets.

**TABLE 6** Nutrient and energy retention in gilthead seabream juveniles fed the experimental diets for 80 days.

| Retention (% intake) | CTRL | PLANT | PAP | EMERG |
|---|---|---|---|---|
| Protein | 23.1 ± 1.1 $^{b}$ | 28.6 ± 3.2 $^{a}$ | 28.3 ± 1.7 $^{a}$ | 19.7 ± 0.2 $^{b}$ |
| Phosphorus | 41.0 ± 3.0 | 45.7 ± 4.7 | 42.6 ± 6.0 | 38.0 ± 3.8 |
| Energy | 27.1 ± 0.5 $^{c}$ | 31.8 ± 1.0 $^{b}$ | 35.2 ± 0.7 $^{a}$ | 23.8 ± 1.5 $^{d}$ |

Note: Values are means ± standard deviation ($n$ = 3). Different superscripts within the same row indicate significant differences ($p < .05$) among diets. Absence of superscripts indicates no significant differences.

**TABLE 7** Apparent digestibility coefficients (ADC) of nutrients and energy of experimental diets.

| ADC (%) | CTRL | PLANT | PAP | EMERG |
|---|---|---|---|---|
| Dry matter | 68.3 ± 0.4 $^{a}$ | 67.4 ± 1.6 $^{a}$ | 69.3 ± 0.8 $^{a}$ | 34.1 ± 2.3 $^{b}$ |
| Protein | 92.8 ± 0.7 $^{ab}$ | 93.8 ± 0.9 $^{a}$ | 90.9 ± 0.4 $^{b}$ | 69.8 ± 1.3 $^{c}$ |
| Lipids | 94.4 ± 0.2 $^{a}$ | 93.4 ± 0.5 $^{a}$ | 94.5 ± 0.9 $^{a}$ | 90.0 ± 0.3 $^{b}$ |
| Phosphorus | 58.8 ± 3.9 $^{b}$ | 68.4 ± 0.4 $^{a}$ | 57.2 ± 2.6 $^{b}$ | 62.4 ± 4.7 $^{ab}$ |
| Energy | 84.1 ± 1.0 $^{a}$ | 83.4 ± 1.3 $^{a}$ | 85.0 ± 0.6 $^{a}$ | 64.7 ± 0.9 $^{b}$ |

Note: Values are means ± standard deviation ($n$ = 3). Different superscripts within the same row indicate significant differences ($p < .05$) among diets.

**TABLE 8** Apparent digestibility coefficients (ADC) of amino acids of experimental diets.

| ADC (%) | CTRL | PLANT | PAP | EMERG |
|---|---|---|---|---|
| Arg | 89.3 ± 0.3 [b] | 94.0 ± 0.3 [a] | 93.9 ± 0.2 [a] | 81.6 ± 0.5 [c] |
| His | 94.9 ± 0.1 [a] | 95.3 ± 0.3 [a] | 94.7 ± 0.1 [a] | 84.8 ± 0.6 [b] |
| Lys | 94.8 ± 0.1 [a] | 93.3 ± 0.1 [b] | 95.0 ± 0.1 [a] | 82.3 ± 0.6 [c] |
| Thr | 88.1 ± 0.4 [c] | 89.5 ± 0.2 [b] | 90.7 ± 0.2 [a] | 76.2 ± 0.8 [d] |
| Ile | 90.9 ± 0.3 [a] | 85.6 ± 0.3 [c] | 88.9 ± 0.1 [b] | 74.0 ± 1.1 [d] |
| Leu | 91.0 ± 0.3 [a] | 86.6 ± 0.5 [b] | 90.3 ± 0.2 [a] | 69.1 ± 1.0 [c] |
| Val | 86.8 ± 0.4 [ab] | 83.7 ± 0.5 [bc] | 89.6 ± 0.3 [a] | 61.3 ± 3.7 [c] |
| Met | 96.4 ± 0.1 [a] | 95.8 ± 0.2 [b] | 95.5 ± 0.1 [b] | 93.5 ± 0.3 [c] |
| Phe | 90.3 ± 0.4 [a] | 86.1 ± 0.3 [c] | 88.6 ± 0.3 [b] | 76.3 ± 1.3 [d] |
| Cys | 94.7 ± 0.2 [b] | 95.7 ± 0.1 [a] | 96.3 ± 0.0 [a] | 95.7 ± 0.4 [a] |
| Tyr | 90.4 ± 0.3 [c] | 91.3 ± 0.1 [b] | 92.4 ± 0.1 [a] | 78.0 ± 0.7 [d] |
| Asx | 91.2 ± 0.2 [a] | 85.6 ± 0.6 [c] | 88.5 ± 0.4 [b] | 79.9 ± 0.7 [d] |
| Glx | 91.2 ± 0.3 [c] | 93.6 ± 0.2 [a] | 92.6 ± 0.2 [b] | 84.6 ± 0.7 [d] |
| Ala | 91.2 ± 0.2 [a] | 86.5 ± 0.6 [b] | 91.9 ± 0.2 [a] | 68.1 ± 0.7 [c] |
| Gly | 87.8 ± 0.1 [b] | 83.5 ± 0.5 [c] | 92.0 ± 0.2 [a] | 70.8 ± 1.2 [d] |
| Pro | 91.0 ± 0.4 [bc] | 92.5 ± 0.2 [ab] | 94.5 ± 0.0 [a] | 72.8 ± 1.3 [c] |
| Ser | 91.0 ± 0.3 [a] | 87.5 ± 0.5 [b] | 90.7 ± 0.3 [a] | 78.6 ± 1.0 [c] |
| Tau | 97.2 ± 0.3 [a] | 82.6 ± 0.3 [bc] | 95.6 ± 0.1 [ab] | 78.7 ± 1.9 [c] |

Note: Values are means ± standard deviation (*n* = 3). Different superscripts within the same row indicate significant differences ($p < .05$) among diets.

**FIGURE LEGENDS:**

**FIGURE 1** Daily nitrogen (N) balance in gilthead seabream juveniles fed the experimental diets for 80 days. Values are means ± standard deviation (*n* = 3). Different letters within bars indicate significant differences ($p < .05$) among diets. Absence of letters indicate no significant differences.

**FIGURE 2** Daily phosphorus (P) balance in gilthead seabream juveniles fed the experimental diets for 80 days. Values are means ± standard deviation (*n* = 3). Different letters within bars indicate significant differences ($p < .05$) among diets. Absence of letters indicate no significant differences.

**FIGURE 3** Transepithelial electrical resistance (Rt; Upper panel) and short-circuit current (Isc; Lower panel) in the anterior intestine of gilthead seabream juveniles fed the experimental diets for 12 weeks. Values are means ± standard deviation (*n* = 7-8). Different letters indicate significant differences ($p < .05$) among diets. Absence of letters indicate no significant differences.

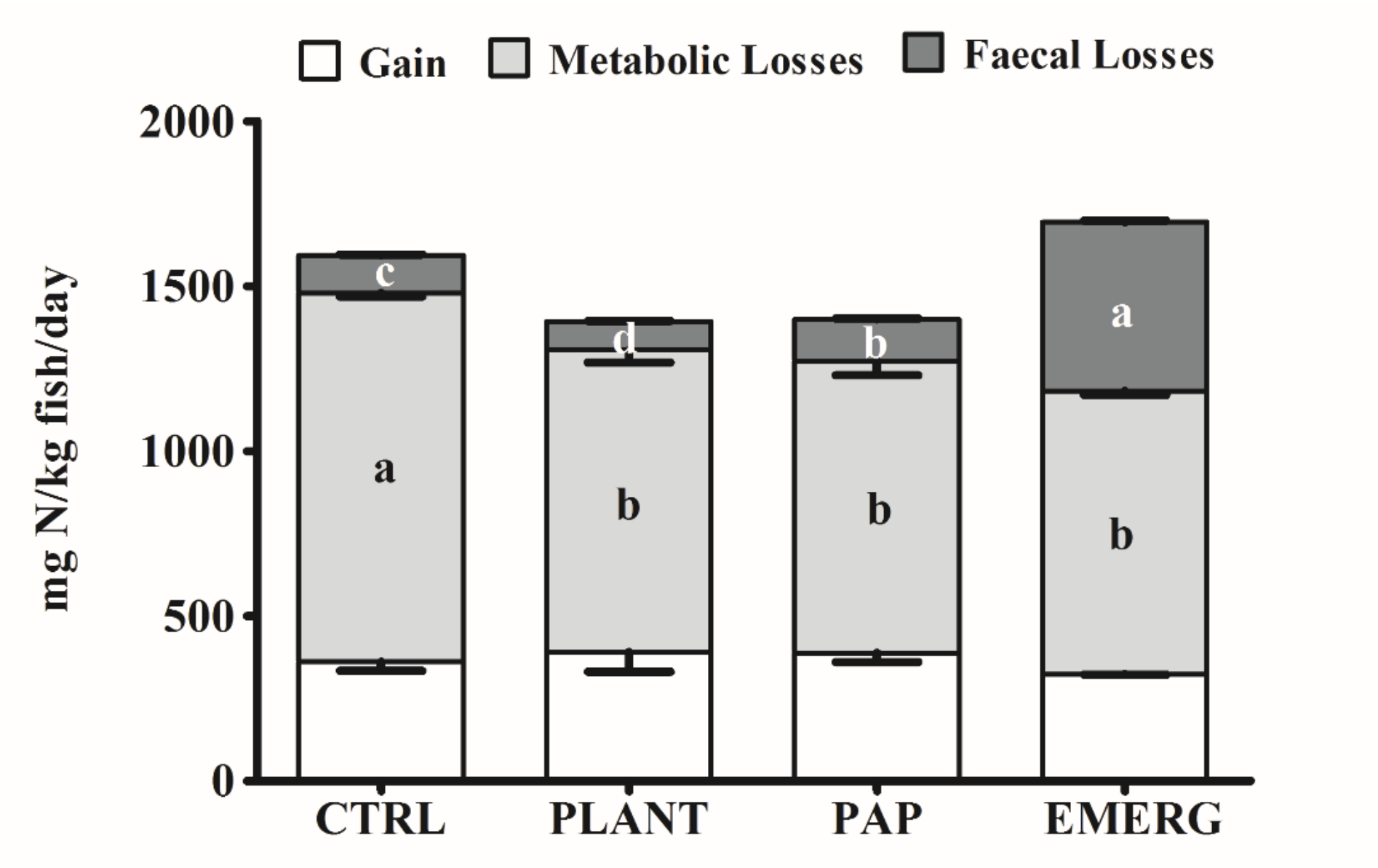

**Figure 1:** Aragão et al.

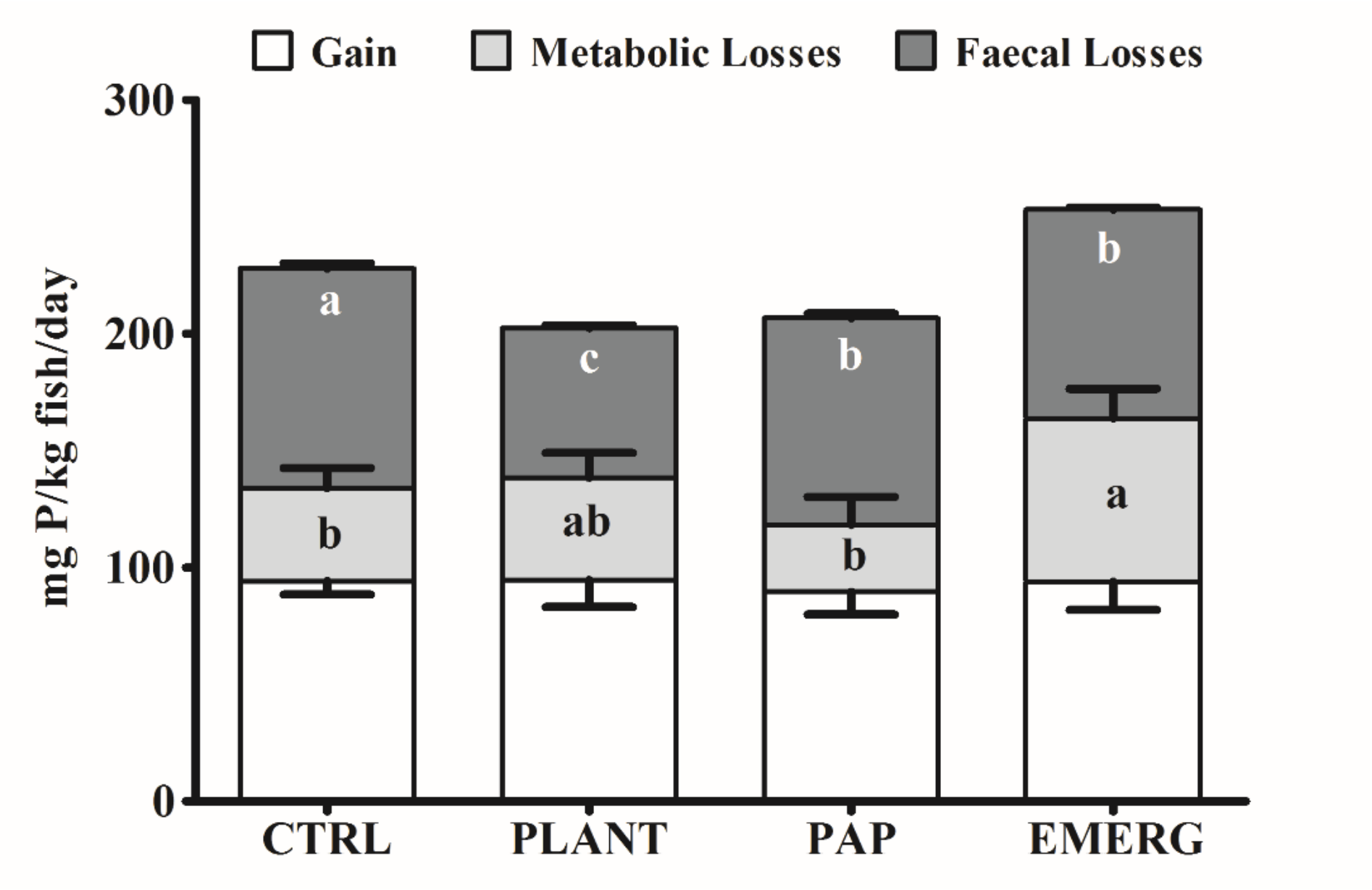

**Figure 2:** Aragão et al.

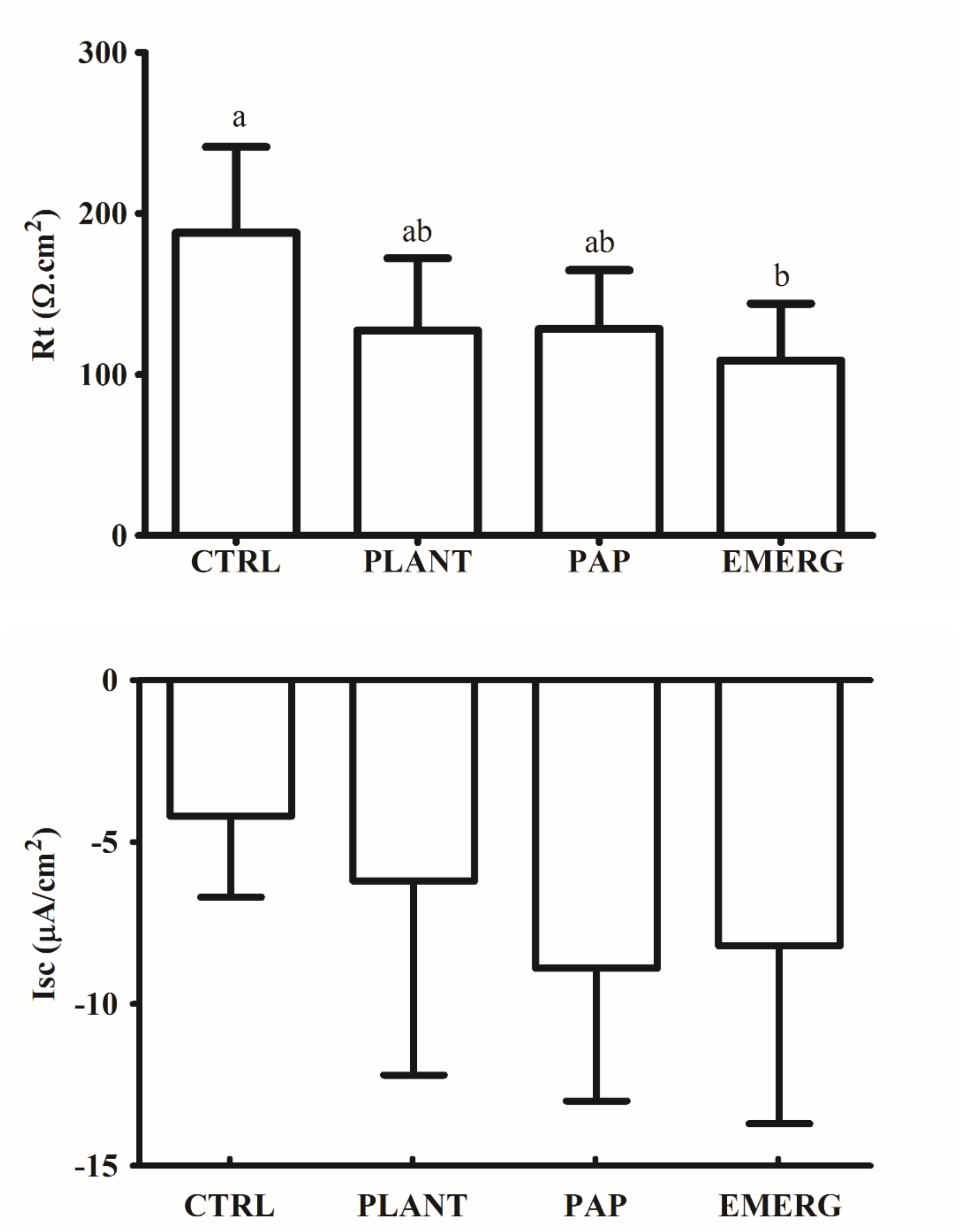

**Figure 3:** Aragão et al.